\documentclass[twocolumn,showpacs,preprintnumbers,amsmath,amssymb]{revtex4}

\usepackage{graphicx}
\usepackage{epstopdf}
\usepackage{dcolumn}
\usepackage{bm}
\usepackage{multirow}
\usepackage{natbib}

\begin{document}
\preprint{White et al}

\title{ Melting and phase change for laser-shocked iron}

\author{S White}
\email{swhite06@qub.ac.uk}
\affiliation{Centre for Plasma Physics, School of Mathematics and Physics, Queen's University Belfast, University Road, Belfast BT7 1NN, UK}
\author{B Kettle}
\affiliation{Centre for Plasma Physics, School of Mathematics and Physics, Queen's University Belfast, University Road, Belfast BT7 1NN, UK}
\author{CLS Lewis}
\affiliation{Centre for Plasma Physics, School of Mathematics and Physics, Queen's University Belfast, University Road, Belfast BT7 1NN, UK}
\author{D Riley}
\affiliation{Centre for Plasma Physics, School of Mathematics and Physics, Queen's University Belfast, University Road, Belfast BT7 1NN, UK}
\author{J Vorberger}
\affiliation{Helmholtz-Zentrum Dresden-Rossendorf, 01328 Dresden, Germany}
\author{SH Glenzer}
\affiliation{SLAC National Accelerator Laboratory, Menlo Park, CA 94025, USA}
\author{E Gamboa}
\affiliation{SLAC National Accelerator Laboratory, Menlo Park, CA 94025, USA}
\author{B Nagler}
\affiliation{SLAC National Accelerator Laboratory, Menlo Park, CA 94025, USA}
\author{F Tavella}
\affiliation{SLAC National Accelerator Laboratory, Menlo Park, CA 94025, USA}
\author{HJ Lee}
\affiliation{SLAC National Accelerator Laboratory, Menlo Park, CA 94025, USA}
\author{CD Murphy}
\affiliation{University of York, Department of Physics, Heslington, York YO10 5DD England}
\author{DO Gericke}
\affiliation{University of Warwick, Department of Physics, Controlled Fusion, Space and Astrophysics, Coventry CV4 7AL, W Midlands, England}

\date{\today}

\begin{abstract}
Using the LCLS facility at the SLAC National Accelerator Laboratory, we have observed X-ray scattering from iron compressed with laser driven shocks to Earth-core like pressures above 400 GPa. The data shows shots where melting is incomplete and we observe hexagonal close packed (hcp) crystal structure at shock compressed densities up to 14.0 g cm$^{-3}$ but no evidence of a double-hexagonal close packed (dhcp) crystal. The observation of a crystalline structure at these densities, where shock heating is expected to be in excess of the equilibrium melt temperature, may indicate superheating of the solid. These results are important for equation of state modelling at high strain rates relevant for impact scenarios and laser-driven shock wave experiments.\end{abstract}

\maketitle

Warm dense matter (WDM) is an intermediate state between plasmas and condensed matter \citep{leerw, ng, koenig}, characterised by strong inter-particle interaction, partial degeneracy and partial ionisation. Such states of matter exist in the interiors of planets and understanding the thermodynamic properties of iron at the Earth core requires detailed knowledge of this state. They are challenging to understand theoretically as methods used in plasma physics for hotter, less dense plasmas and condensed matter physics for high density, cooler states are not suitable as the thermal energy is of the order of the potential energy and no small expansion coefficient exists.

Due to its occurrence in the cores of Earth-like planets, iron is of particular interest and both static compression and shock wave experiments have been used to explore the equation of state and melting at high pressure \citep{boehler,williams,nguyen,brownmcqueen,bancroft,jamieson,kalantar,brown,brown2,dubrovinsky,yoo2,vocaldo,belonshonko}. It is expected e.g. \citep{nguyen,brownmcqueen} that shock melting begins at about 220 GPa and is complete by 280 GPa. A phase change from bcc to hcp occurs in both static and shock compression at a relatively modest pressure of $\sim$13 GPa \citep{bancroft,jamieson,kalantar}. Further phase changes have been discussed for both shock \citep{brown,brown2} and static compression \citep{dubrovinsky,yoo2}. A transition from hcp to dhcp has been reported \citep{yoo2,vocaldo}, as well as stability of the bcc structure at Earth core conditions \citep{belonshonko} and stability of the $\gamma$ (fcc) phase for highly compressed Fe. e.g. \citep{stixrude,pickard}. 

\begin{figure}[htpb]
\centering
  \includegraphics[width=7.5cm]{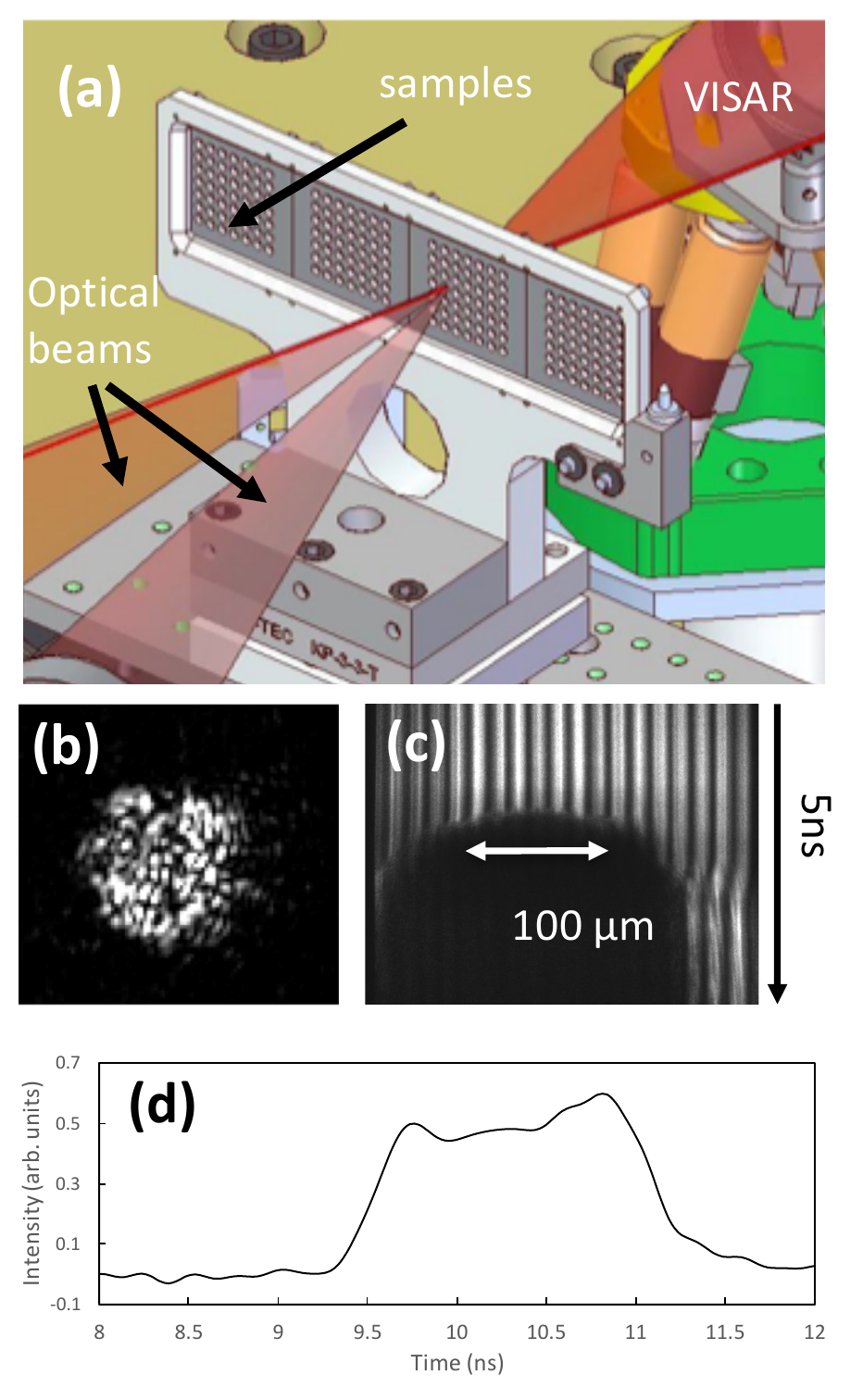} \\[1pt]
 \caption{(Colour online) (a) Schematic of the experimental arrangement. The LCLS beam (solid red line) passed through the samples at an angle of 24$^{\circ}$ to the target normal with 20 $\mu$m diameter. The optical beams were incident on the target at an angle of 16$^{\circ}$ either side of the target normal. (b) Focal spot image of the beams with the 100 $\mu$m phase plate (c) VISAR data showing breakout when using the 100 $\mu$m phase plate. The central 100 $\mu$m region is overall relatively flat. (d) Oscilloscope trace of the optical pulse shape. }\label{experiment}
\end{figure}

We report X-ray scattering measurements from samples of warm dense iron created using laser driven shock compression at the MEC end-station of the LCLS X-ray free electron laser \citep{glenzer,nagler}. The target samples were iron foils of 10.2$\pm$0.3 $\mu$m thickness and coated with 5.0 $\pm$0.1 $\mu$m of CH on one side. The two optical laser beams of the MEC end-station (527 nm wavelength) were focused with the use of random phase plates to a focal spot of either 100 or 50 $\mu$m diameter onto the CH coated side of the samples. The optical pulse shape, as can be seen in Fig. \ref{experiment} (d), rose in about 0.5 ns with a FWHM of 1.6 ns peaking at intensity of 2$\times$10$^{13}$ Wcm$^{-2}$ (100 $\mu$m spot shot) before falling off over 0.5 ns. The LCLS beam was focused with a Be lens to a spot of 20 $\mu$m diameter and centred on the optical focal spot.

A VISAR system monitored the shock break-out from the rear of the samples. We see in Fig. \ref{experiment}(c) the shock affected area is wider than the nominal 100 $\mu$m because with phase plates 20$\%$ of the energy is diffracted outside of the central spot and there is lateral spread of the beam in the CH plasma and shock spreading, although the total target thickness is six times less that the nominal spot size. The shock break-out was accompanied by a drop in reflectivity associated with the heating and decompression of the rear surface. The principal diagnostic of the scattered X-rays was the 560K CSPAD detector \citep{herrmann}. Figure 2 shows raw data images from the CSPAD for two shots. The bcc diffraction lines are caused by scattering from cold foil surrounding the heated sample region. In situ testing showed that this is caused when the beryllium lens scatters a small portion of the beam ($\sim$ 1$\%$) into a 'halo' around the central bright spot. This gives clear angular calibration features but plays no role in the interpretation of our data otherwise. Characterisation of the cold targets indicates that they are polycrystalline with crystallites of $\sim$micron size and no measurable porosity.

\begin{figure}[htbp]
\centering
  \includegraphics[width=8.5cm]{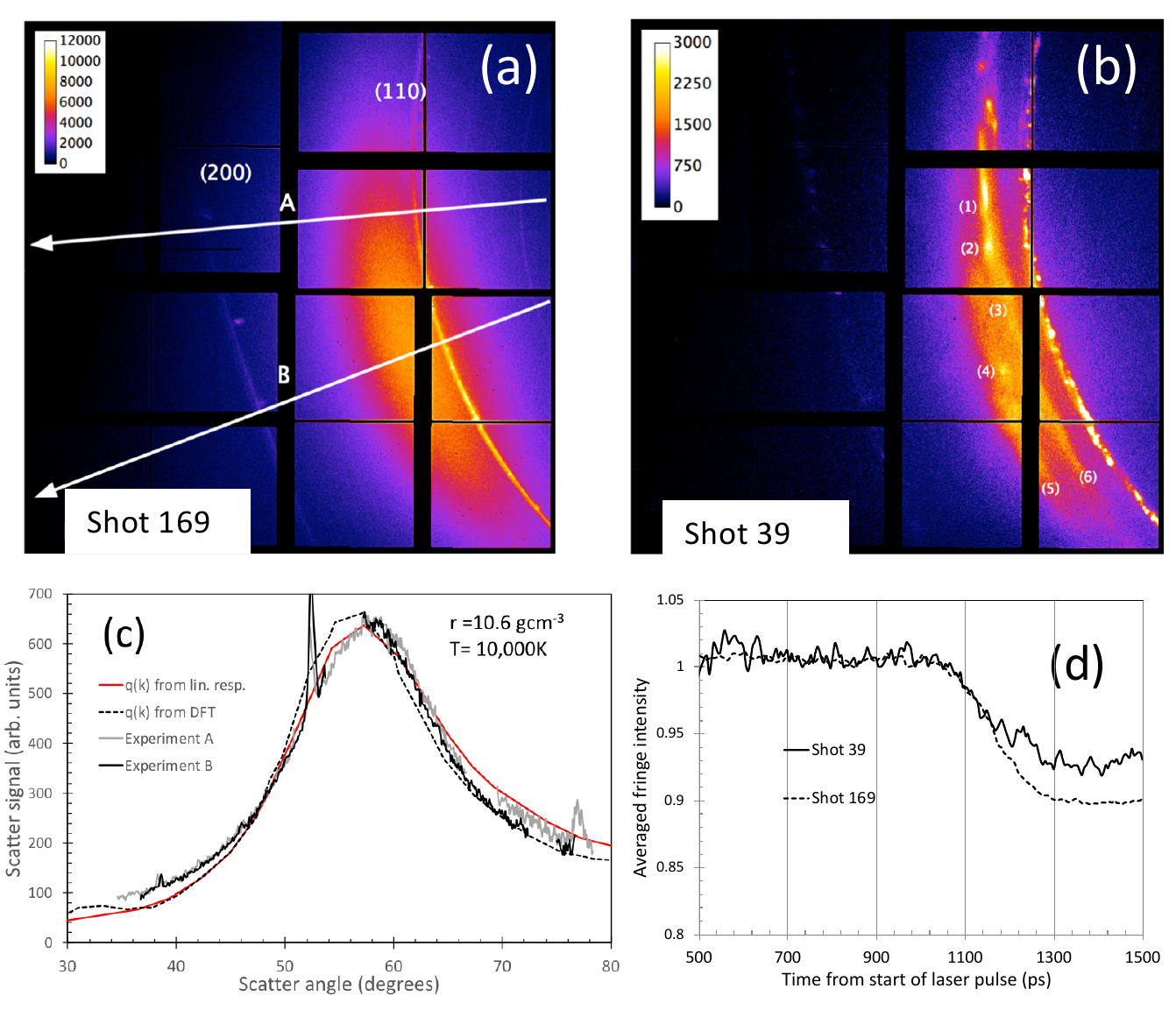} \\[1pt]
 \caption{(a) Raw diffraction data for shock melted iron. The probe time was $\sim$40 ps before shock break-out. (b) Data for a shot similar to that in (a) with probe time 170 ps ahead of shock break-out. The features marked 1 to 3 and 6 are identified as hcp 002 reflections whilst 4 and 5 are hcp 101. The 002 reflection is underneath the cold bcc 110 reflection in this shot. (c) Two lineouts of the melted shot in (a) showing good consistency as a check on the image processing. The data is fitted with a DFT simulation with the electron-ion correlation term calculated in two ways as described in the text. (d) Comparison of the VISAR data for the two shots indicates that it is not possible to distinguish within error bars a difference in the shock speed in the two shots}\label{rawdata}
\end{figure}

We see complete melting of the iron in Fig. \ref{rawdata}(a) with the expected liquid-type diffraction feature which can be described with density functional molecular dynamics (DFT-MD). In Fig.\ref{rawdata}(b) melting has not occurred and we see strong evidence of the hcp crystalline phase. The densities assigned to individual hcp diffraction features range from 10.7 to 11.1 gcm$^{-3}$. Fig.\ref{rawdata}(c) shows a line-out of the liquid diffraction feature from Fig.\ref{rawdata}(a). We have modelled the data using ab initio DFT-MD simulations. The structure may be understood as that of a strongly coupled one component plasma with ion charge state $Z=8$ and degenerate Yukawa type screening of the ion-ion interaction (see supplementary material). Fitting to the experimental data gives a density of 10.6 gcm$^{-3}$ $\pm$ 0.3 gcm$^{-3}$ with a temperature of 10$^{4}$ $\pm$ 0.2$\times$10$^{4}$K. The error bars are estimated by performing least squares fits to a series of DFT-MD simulations, adjusting the scaling constant between experiment and simulation in each shot to minimise the least squares sum. This is similar to the density range seen from the crystalline features in the un-melted shot, Fig.\ref{rawdata}(b). In Fig.\ref{rawdata}(d) we see normalised average fringe intensities for both shots. The shock break-out time is very similar, although, post break-out, there is a difference in the fringe intensity. This may indicate a difference in the expanded plasma conditions linked to the melting/non melting conditions. Analysis of both VISAR channels indicates for the melted shot we have break-out at 1.04 $\pm$ 0.02 ns after the start of the shock driving beam. The probe time was set to 1.0 ns, so just prior to shock break-out. For the un-melted shot, break-out was 1.07 $\pm$ 0.02 ns and probing was at 0.9 ns. 

As noted, scatter from the Be lens causes the smooth bcc diffraction in Fig.\ref{rawdata}(a). In Fig.\ref{rawdata}(b), where we probe prior to shock break-out, we also see bright spots due to diffraction from crystallites contained with the small volume of unshocked iron just ahead of the shock (~1$\mu$m thick). This feature is seen when we probe before the shock break-out time and is not seen on shots where we probe after break out. 

\begin{figure}[htbp]
\centering
  \includegraphics[width=7.5cm]{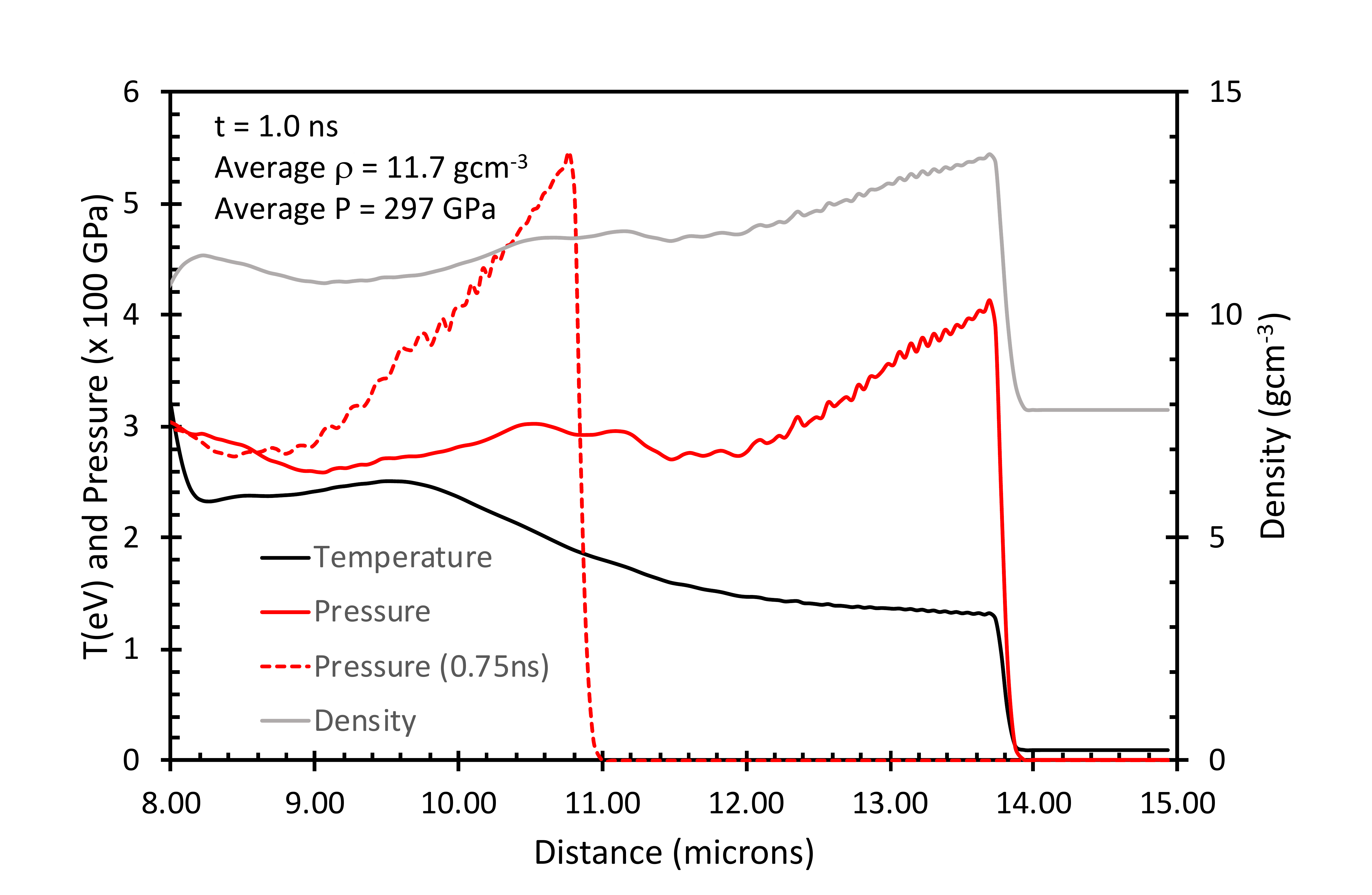} \\[1pt]
 \caption{(Colour online) Hydrodynamic simulation of shot 169 using the Hyades code with the measured pulse shape and energy. The code uses multi-group diffusion for radiation transport in 35 groups logarithmically spaced from 0.01 to 15 keV. The ionisation model is Thomas-Fermi and SESAME equation of state 2140 for Fe is used. We have also plotted the pressure profile at an earlier time (0.75 ns) to illustrate the fact that the shock pressure is decaying and that the whole of the target is subjected to shock pressure and temperature well in excess of what is expected to be required for melting. }\label{hyades}
\end{figure}

We have performed a hydrodynamic simulation of shot 169 (Fig. \ref{hyades}) with the HYADES code \citep{larsen} using the SESAME equation of state 2140 for iron \citep{sesame}. The shock front has a density $\sim$13.5 gcm$^{-3}$ and velocity ~11.5 km/s. These parameters are in good agreement with published measurements for the 400 GPa shock front pressure \citep{brown}. The simulated averaged density in the iron is 11.7 gcm$^{-3}$, about 10$\%$ higher than our DFT fit. The shock front temperature is 1.3 eV (15,000K) which is higher than our DFT fit and the latter therefore has a lower pressure of approximately 165 GPa. Both temperatures are well above the expected equilibrium melt temperature for this pressure range \citep{anzellini, bouchet, medvedev}. The simulated temperature is higher behind the shock front as the shock at this point is decaying from a peak simulated pressure close to 10 Mbar which occurs as the shock enters the iron at around 400ps after the start of the optical laser pulse. This is an important point as we probe the whole volume of the iron and due to the decaying shock expect our conditions at the time of probing to be off the Hugoniot away from the region where the equation of state is tested against experiment.

\begin{figure*}[htbp]
\centering
  \includegraphics[width=15cm]{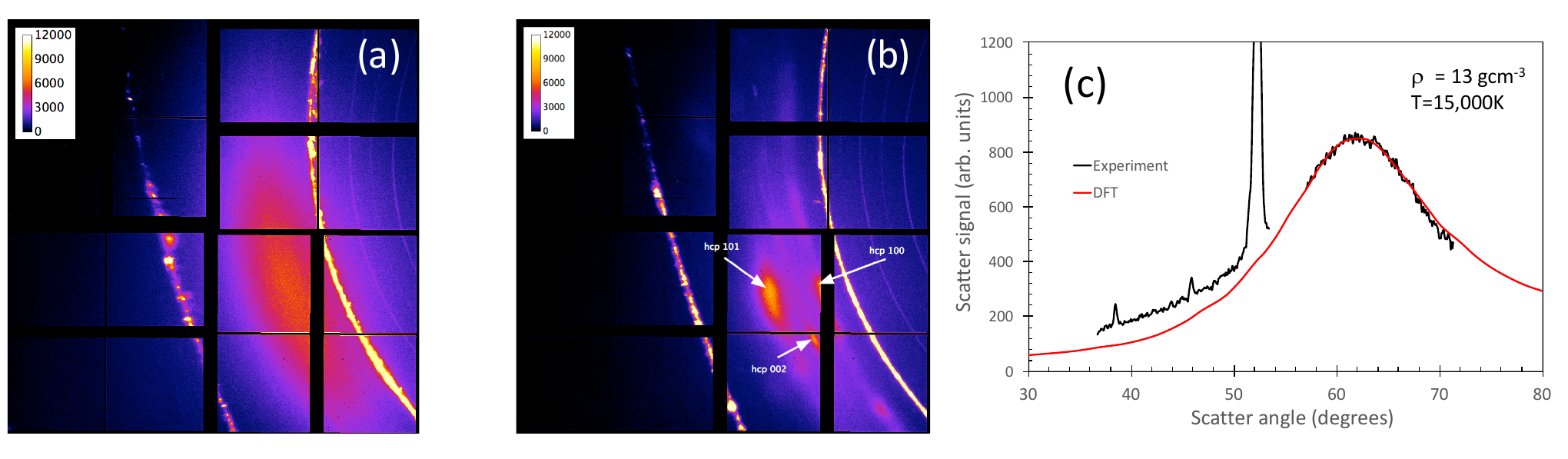} \\[1pt]
 \caption{(Colour online) Data for  50 $\mu$m optical focal spot. (a) Shock breakout was 1.0 ns and probing at 0.8 ns. We see a liquid Fe sample that can be fitted to 12.5 gcm$^{-3}$. (b) The shock break-out is 1.1 ns and probing is at 0.5 ns. There are clear hcp diffraction features noted. The Bragg positions of the various features indicate density ranging from 12.7 to 14.0 gcm$^{-3}$ with a mean and standard deviation of 13.4 $\pm$ 0.5 gcm$^{-3}$. (c) DFT simulation of the data in (a) showing a fit to 13.0 gcm$^{-3}$ and 1.5$\times $10$^{4}$K. The fit is not so close as for the lower density shot. This is most likely due to the smaller focal spot, meaning that non-uniformity in the conditions probed start to become more significant. }\label{rawdata50}
\end{figure*}

Data taken with 50 $\mu$m focal spots had only slightly faster shock speeds due to stronger non-planar behaviour, mostly as a result of the much longer scale-length of the CH plasma but also the smaller ratio of spot-size to target thickness. However, as seen in Fig. \ref{rawdata50}, we collected consistent clear data showing higher density for shots both with melted and with crystalline iron. For unmelted iron we see the hcp phase with features indicating densities between 12.7 to 14.0 gcm$^{-3}$ and the melted shot can be fitted using DFT-MD to a density of 13.0 g cm$^{-3}$ $\pm$ 0.5 g cm$^{-3}$. The temperature best fit is 1.5$^{+0.3}_{-0.1}$ $\times$10$^{4}$K. The unmelted shot is for early probing 0.6 ns before shock-break out. We have similar data for shots where we probe close to shock-breakout but the data is slightly saturated and is shown in the supplementary material. The fit to the melted shot is not as good as for the data in figure 2, since, in addition to the spatial variation in the shock propagation direction, seen in the hydrodynamic simulation, there are stronger lateral gradients.

The shock break-out times observed are inconsistent with shock speeds in Fe below 10 kms$^{-1}$ and thus the average shock pressures are above 300 GPa in all presented shots and we certainly expect melting. An explanation for our data may be the experimentally observed phenomenon of super-heating \citep{boness, mei, luo1, luo2, millot, luo3}. The degree to which a solid may be superheated above the equilibrium melting temperature \citep{luo1} is a function of the energy barrier required for nucleation. This is dependent on the surface tension between the liquid and solid phases and the latent heat of fusion and is a slowly varying quantity with heating rate. The calculated maximum degree of superheating varies for different elements. Luo and Ahrens \citep{luo1} show that a maximum superheating of around 25$\%$ above the equilibrium melt temperature for iron is expected at heating rates of $\sim$10$^{12}$ Ks$^{-1}$, a rate we exceed by an order of magnitude. This result can be obtained using parameters (including equilibrium melting temperature) for iron under ambient conditions but, relevant to our data, Luo and Ahrens give arguments to show that for $\sim$ 100 GPa pressures the value of the energy barrier to nucleation varies only slowly as a function of pressure \citep{luo3}. In addition, the rate of superheating is also very weakly dependent on the heating rate. Thus, under the theory of Luo and Ahrens, we can expect to have a maximum superheating of approximately 25$\%$, and a melt temperature $\sim$ 9000 K based on an equilibrium melt temperature of $\sim$ 7000K for our shock compressed pressure \citep{anzellini,bouchet,medvedev}. This is broadly consistent with the temperature seen in the DFT fits to the melted shot data for figure 2 and hence superheating can explain the fact that we see melting on some shots and on other shots with similar shock parameters see crystalline features at similar compression. 

Evidence of super-heating has been reported previously for shocked Fe \citep{luo3} and other materials \citep{boness}, based on the emission glow of a sample either during shock propagation for transparent materials or after shock break-out for opaque materials. In this work we do not have optical emission data but have direct structural evidence from the bulk of the sample that shows us not only that there is a crystalline structure but that it is hcp which is an expected phase change for low shock pressure and not dhcp which has been reported as a solid-solid phase change \citep{yoo2, vocaldo} closer to 200 GPa. As seen in figure 4, we also see clear hcp features on shots where the corresponding melted shot (i.e. similar shock break-out time) has a fit temperature well in excess of 10$^{4}$ K.

Rethfeld et al \citep{rethfeld}  calculated the homogeneous nucleation time for different materials using values for the surface tension between the liquid and solid phases, heat capacity and latent heat of melting for iron under ambient conditions. These data allow us to estimate that the nucleation time, for 25$\%$  superheating, is above the nanosecond level of our shock driving pulses but there is a very rapid dependence on the level of heating.

In our data, we see shock compressed iron at averaged densities up to 14.0 gcm$^{-3}$ where the sample is in the hcp crystalline phase. We believe this is clear evidence of superheating for the following reasons. Firstly, longer timescale ($\mu$s) flyer plate measurements of the Hugoniot for Fe \citep{brown} give a shock speed of 10 kms$^{-1}$ for a pressure of 300 GPa which is above the expected upper limit for melting \citep{nguyen}. In this study, at nanosecond time scales, the shock break out times for both melted and unmelted case are consistent with modelling that shows a shock velocity of $\sim$ 11.5 kms$^{-1}$ and pressure $\sim$ 400 GPa on shock exit.

The hydro-simulation in Fig. \ref{hyades} indicates a higher temperature behind the shock front. This is due to decay of the shock coupled to the shock reverberation expected due to the CH/Fe interface \citep{swift}. The DFT fitted temperature for melted Fe, is lower than the hydro-simulation indicates and this may be due to assumptions made in the EOS away from the principal Hugoniot where it is tested against experiment. The DFT temperature of 10$^{4}$ K is sufficiently close to the range of the calculated static melt temperature to be consistent with superheating theory if we assume a similar temperature for the un-melted shot. For the higher temperatures of Fig. \ref{rawdata50}, we can point to the timescale as limiting the rate at which melting can be achieved. The rate is implicit in the calculations of Luo and Ahrens \citep{luo1} and the explicit calculations for nucleation time from Rethfeld et al \citep{rethfeld} can be of order of the pulse duration.

In summary, we have observed X-ray scattering from iron under shock compressed conditions where completion of melting would be expected. The data has two significant outcomes. Firstly, it suggests that superheating might be a key phenomenon in our experiment and we have observed this not through observing shock break-out emission and inferring a temperature as in previous work [29] but by direct observation of the shock compressed crystal structure. Secondly, we have seen no evidence of dhcp or another solid crystalline phase. Such a further phase change has been reported by others for static compressions and for longer timescales in shock experiments. Our results indicate that there is a limit on rapidity for this phase change that is important to consider in the implementation of equations of state for hydrodynamic modelling of laser-driven shocks  and high pressure impact scenarios on nanosecond timescales. 

section{Acknowledgments}
This work was supported by UK Engineering and Physical Sciences Research Council grant No. EP/K009591/1. Supplementary data are available via www.qub.ac.uk/research. Use of the Linac Coherent Light Source (LCLS), SLAC National Accelerator Laboratory, is supported by the U.S. Department of Energy, Office of Science, Office of Basic Energy Sciences under Contract No.  DE-AC02-76SF00515. The MEC instrument is supported by the U.S. Department of Energy, Office of Science, Office of Fusion Energy Sciences under contract No. SF00515.

\end{document}